\documentclass[letterpaper,twoside,twocolumn,english,final, prb, aps, superscriptaddress]{revtex4-1}
\usepackage[T1]{fontenc}
\usepackage[latin9]{inputenc}
\usepackage{textcomp}
\usepackage{amsmath}
\usepackage{amssymb}
\usepackage{graphicx}

\makeatletter

\pdfpageheight\paperheight
\pdfpagewidth\paperwidth

\usepackage{pslatex}
\usepackage{bm}

\makeatother

\usepackage{babel}
\begin{document}
\title{Structural phase transition of two-dimensional monolayer SnTe from
artificial neural network}
\author{Jiale Zhang}
\affiliation{School of Microelectronics \& State Key Laboratory for Mechanical
Behavior of Materials, Xi'an Jiaotong University, Xi'an 710049, China}
\author{Danni Wei}
\affiliation{School of Microelectronics \& State Key Laboratory for Mechanical
Behavior of Materials, Xi'an Jiaotong University, Xi'an 710049, China}
\author{Feng Zhang}
\affiliation{School of Microelectronics \& State Key Laboratory for Mechanical
Behavior of Materials, Xi'an Jiaotong University, Xi'an 710049, China}
\author{Xi Chen}
\affiliation{Department of Applied Physics, Aalto University, Espoo 00076, Finland}
\author{Dawei Wang}
\email{dawei.wang@xjtu.edu.cn}

\affiliation{School of Microelectronics \& State Key Laboratory for Mechanical
Behavior of Materials, Xi'an Jiaotong University, Xi'an 710049, China}
\date{\today}
\begin{abstract}
As machine learning becomes increasingly important in engineering
and science, it is inevitable that machine learning techniques will
be applied to the investigation of materials, and in particular the
structural phase transitions common in ferroelectric materials. Here,
we build and train an artificial neural network to accurately predict
the energy change associated with atom displacements and use the trained
artificial neural network in Monte-Carlo simulations on ferroelectric
materials to investigate their phase transitions. We apply this approach
to two-dimensional monolayer SnTe and show that it can indeed be used
to simulate the phase transitions and predict the transition temperature.
The artificial neural network, when viewed as a universal mathematical
structure, can be readily transferred to the investigation of other
ferroelectric materials when training data generated with \emph{ab
initio} methods are available.
\end{abstract}
\maketitle
Ferroelectric materials, which have spontaneous polarizations that
can be reversed by an external electric field, constitute a group
of functional materials that are important for many applications,
e.g., ultrafast switches, phased-array radar, and dynamic random access
memories \citep{ferroelectrics}. It is reported that SnTe thin films
with a thickness of 1-unit cell (UC, with two layers of atoms) can
have stable spontaneous polarization up to 270K, and 2-UC to 4-UC
SnTe films also have strong ferroelectric properties at room temperature
\citep{SnTe}. The ferroelectricity of this two-dimensional (2D) material
makes it a good candidate for applications in promising devices such
as high-density memory, nano-sensors\citep{2Dferroelectrics1,2Dferroelectrics2}.
To fully understand a ferroelectric material such as SnTe, it is necessary
to know if it can indeed experience structural phase transition(s),
and if so, the phase transition sequence, and temperature is. It is
also important to know the relation between the transition temperature
$T_{C}$ and other factors, such as strain and the number of layers.
Since accurately predicting the phase transition temperature requires
large systems, it is usually not possible with a pure \emph{ab initio}
approach, i.e., the density functional theory (DFT) \citep{DFT}.
To overcome the limitation of DFT's huge computational cost for large
systems, one need to propose empirical formulas to fit the inter-atomic
potential energy (or forces) to DFT results using small systems, which
is then applied to large systems. For more complex systems such as
perovskites, the effective Hamiltonian approach has been a popular
method to investigate their static and dynamic properties \citep{prb6301,DWang1,DWang2,DWang3,DWang4}.
For such first-principles-based methods to work, empirical formulas
or proper effective Hamiltonian (along with the specification of the
dynamic variables and the coefficients of the effective Hamiltonian)
are needed, neither of which is trivial if the aim is to achieve a
general and transferable approach for different systems.

\begin{figure}[h]
\begin{centering}
\includegraphics[width=8cm]{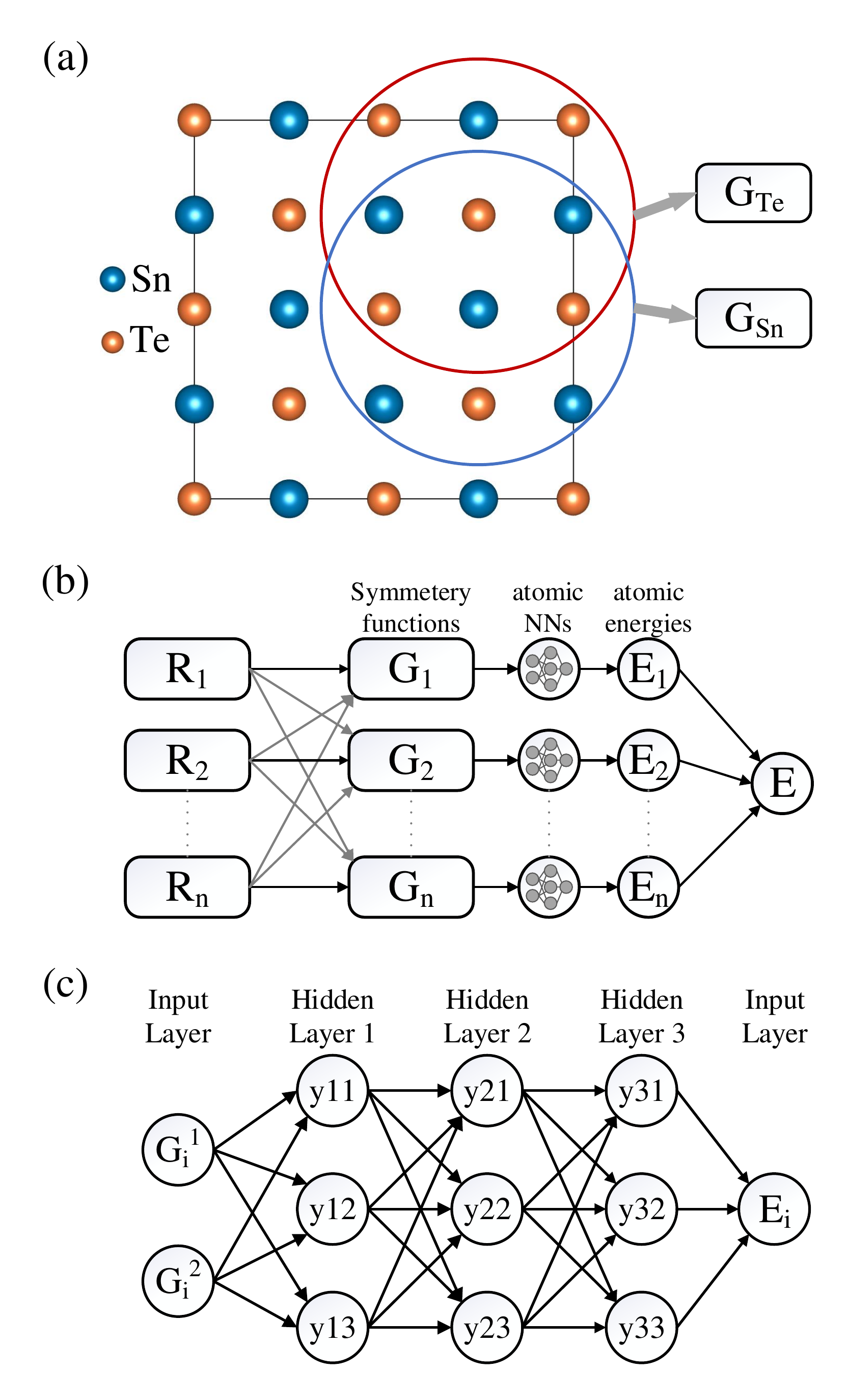}
\par\end{centering}
\caption{\label{fig:1_toplogy}(a) The 2D monolayer SnTe shows alternating
Sn and Te atoms along the $x$ and $y$ axes. The blue (red) box indicates
the chemical environments encircling the Sn (Te) atom at the center;
(b) The topology of the whole structure is built to predict the total
energy, which contains the ANNs that are schematically shown in (c).}
\end{figure}
Recent advances in machine learning (ML) offers an alternative approach
for the construction of the potential-energy surface (PES) by fitting
large data sets from electronic structure calculations with DFT \citep{JorgBehler2016,Deringer2019}
. The ML potential, when viewed as an universal mathematical structure,
has the advantage that it can be used for very different systems with
minimal modification. With such an approach, there is no need to reconstruct
the formulas for different systems, while the accuracy of the PES
remains satisfactory. It is promising that the use of ML techniques
will combine the advantages of both approaches, i.e., the accuracy
of DFT and the efficiency of explicit formulas in obtaining the PES
\citep{tong2020}. While the ML approach has been applied to many
interesting systems \citep{Silicon,sodium,water}, the structural
phase transitions induced by atom displacements, which is important
in ferroelectric materials, have not been dealt with. In this paper,
we develop a neural-network-based approach to treat ferroelectric
systems, especially their phase transitions. We note that such structural
phase changes usually involve atom displacements within unit cells
in crystalline phases, unlike the researches focusing on finding novel
structures in liquid and amorphous phases \citep{ex1,ex3,ex5,Sosso2012}.

In order to construct this framework and show its efficacy, we use
2D monolayer SnTe as the prototypical material to ground this approach.
It is known that SnTe bulk can have a structural phase transition
and the SnTe thin films (of 2 or more layers of atoms) can have a
rather high transition temperature (270 K) \citep{SnTe}, however,
it is unknown if a \emph{monolayer} SnTe can still have a phase transition
when all the atoms are confined to 2D. To address this question, we
adopt the ML approach, build artificial neural networks (ANN) suitable
for the system with two different types of atoms, and take DFT results
as training data for the supervised learning of the ANNs. In this
way, we have successfully constructed the PES (with respect to the
displacement of atoms) and employed it in Monte-Carlo (MC) simulations,
which demonstrate a structural phase transition occurring at \textasciitilde 260\,K.
The approach and the programs we have developed are universal enough
that they can be used to investigate different crystals and shed light
on their structural phase transitions.

The key ingredient in this approach is to build and train an ANN that
can efficiently and accurately predict the energy of a given atomic
configuration. A monolayer SnTe has the simple structure as shown
in Fig. \ref{fig:1_toplogy}(a) where Sn and Te atoms alternate with
each other along both the $\left[100\right]$ and $\left[010\right]$
directions. To fulfill this mission, the whole architecture has the
topology shown in Fig. \ref{fig:1_toplogy} (b), which is similar
to those used by Behler and Parrinell \citep{JorgBehler2016,JorgBehler2014,Behler2007}.
More specifically, the total energy $E$ of the system is the sum
of atomic contributions $E_{i}$,

\begin{equation}
E=\sum_{i}E_{i}
\end{equation}
where $E_{i}$ is the energy imposed on the $i$th atom by its neighboring
environment, which will be determined by the ANN we build. It is realized
that the inter-atomic potential energy decays rapidly with distance
(which is termed the ``nearsightedness''\citep{nearsightedness1,nearsightedness2}),
therefore a cutoff function is usually used to limit the interaction
between atoms to an appropriate range. In this work, we have found
that, considering the interaction up to the 5th nearest neighbors
{[}see Fig. \ref{fig:1_toplogy}(a){]} will produce satisfactory results.
The input to each ANN in Fig. \ref{fig:1_toplogy}(b) is determined
by the coordinates of the $i$th atom and its eight neighboring atoms,
which are encircled in Fig. \ref{fig:1_toplogy}(a) where two situations
(Sn in the center and Te in the center) are indicated. For the 2D
structure, we only consider atom displacements inside the plane, resulting
in an a vector with 18 elements as the input.

To properly and adequately represent the local chemical environment
around an atom, we find that it is imperative to use atom-centered
symmetry functions $G$ as descriptors, which are a series of functions
of atom positions \citep{PRB2013}. As indicated in Ref. \onlinecite{JorgBehler2016},
the number of symmetry functions describing a given structure should
be greater than the degrees of freedom of the described system so
that all information is fully recorded. In Fig. \ref{fig:1_toplogy}(b),
the column named ``atomic NNs'' contains identical ANNs that takes
the chemical environment of an atom, which are encoded in $G$s as
shown by the column named ``symmetry functions'', as input and outputs
the energy$E_{i}$ indicated by the column named ``atom energies''.
Finally, $E_{i}$ is summed to give the total energy $E$.

The core components of the whole structure are the ANNs as shown schematically
in Fig. \ref{fig:1_toplogy}(c), which can be built with much freedom.
For instance, we can use a simple neural network with back-propagation
\citep{BP} or some deep neural networks \citep{DNN}. Here, given
the relatively simple chemical environments, we constructed an ANN
with three hidden layers, each layer containing 40 nodes. In addition,
since there are two types of atoms (Sn and Te) in the system, two
separate ANNs, which have the same structure but different weights
inside its nodes, were established to calculate the two types energies,
$E_{\textrm{Sn}}$ and $E_{\textrm{Te}}$ imposing on Sn and Te atoms,
respectively.

\begin{figure}
\begin{centering}
\includegraphics[width=6cm]{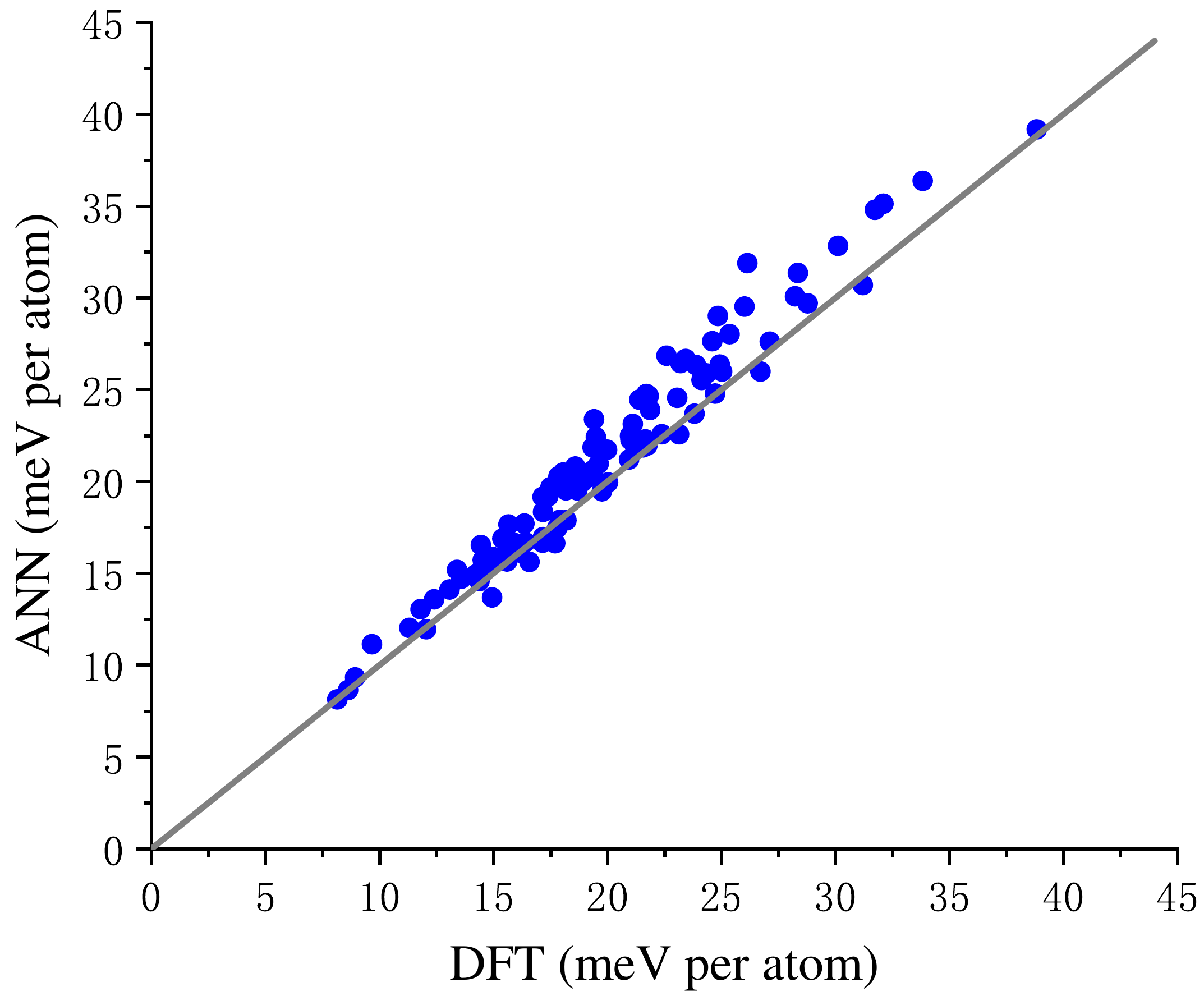}
\par\end{centering}
\caption{\label{fig:error}Comparison of the energy predicted by the trained
ANN and generated by DFT (using GPAW) where for each data point, its
$x$-axis value is from GPAW while its $y$-axis value is the ANN
prediction. }
\end{figure}
\begin{figure}[h]
\begin{centering}
\includegraphics[width=4cm]{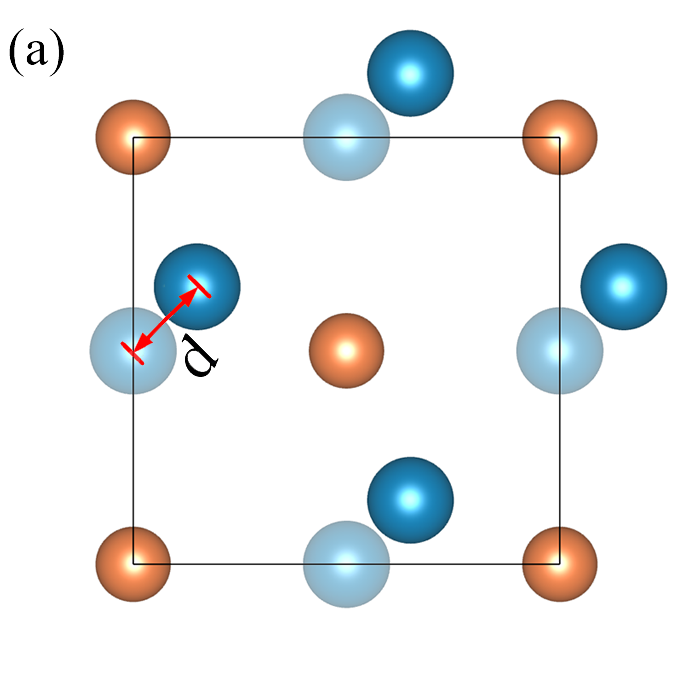}\includegraphics[width=4cm]{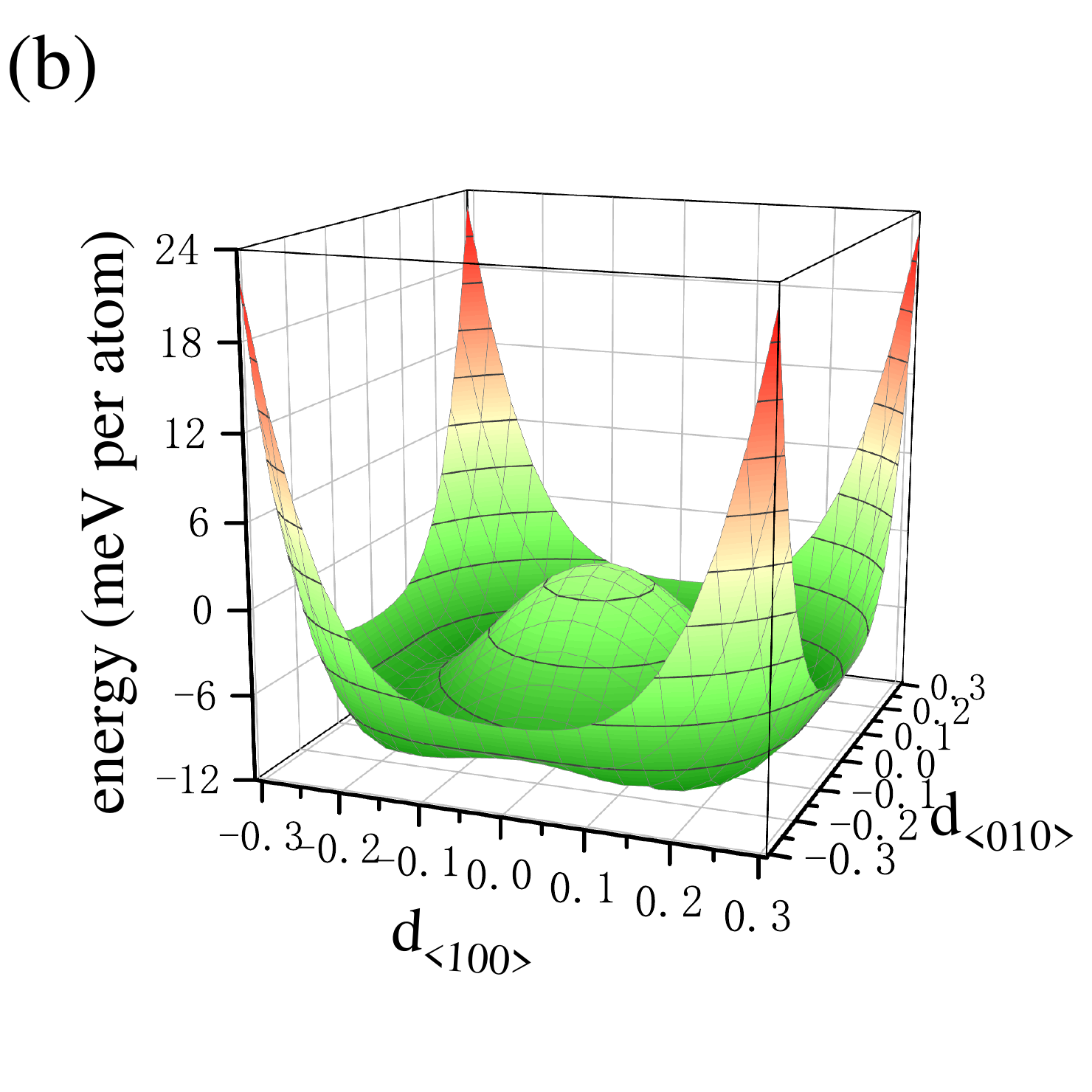}
\par\end{centering}
\begin{centering}
\includegraphics[width=8cm]{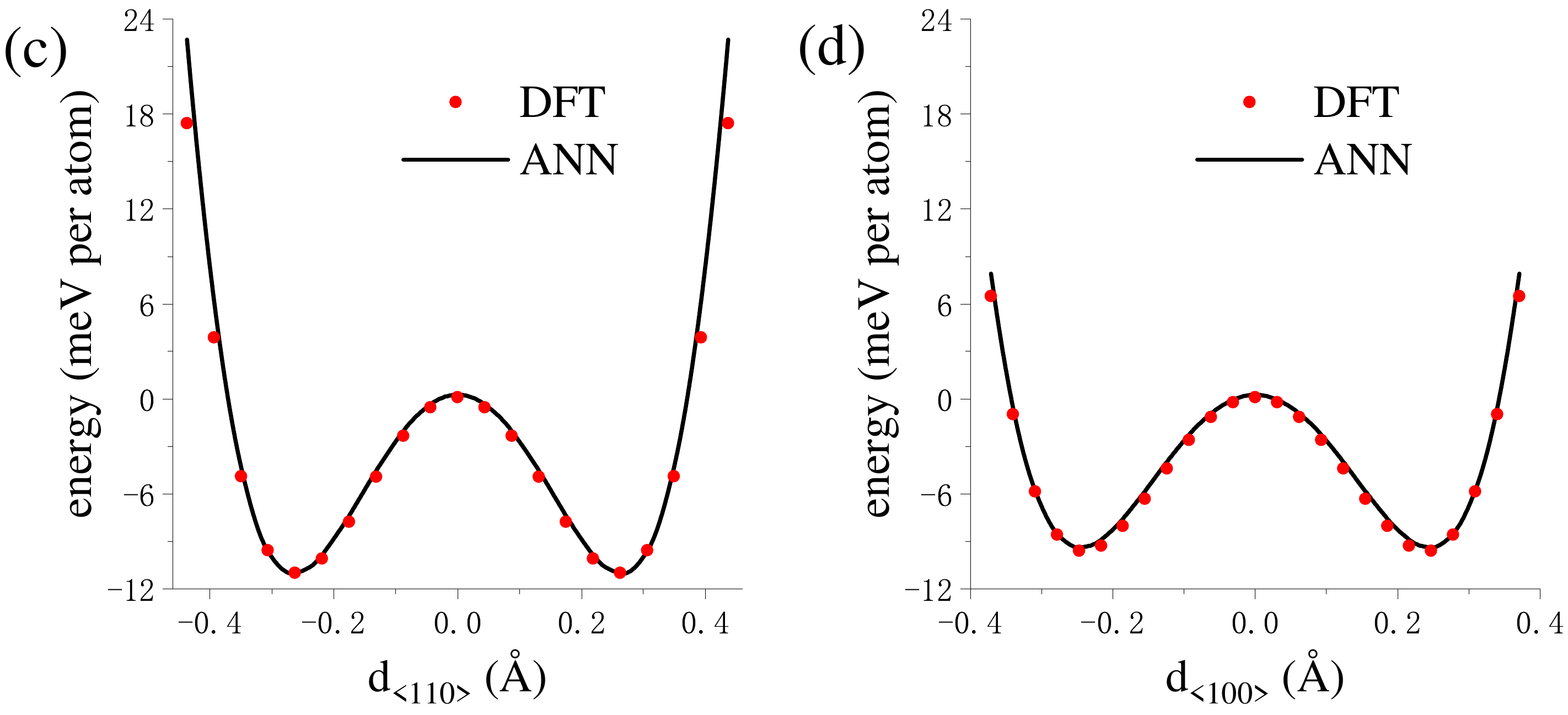}
\par\end{centering}
\caption{\label{fig:result}(a) All the Sn atoms in a unit cell move synchronously
to a distance $d$; (b) The PES is generated by the ANN, showing multiple
energy minimums; (c) and (d) compares the values predicted by the
ANN along the $\left\langle 110\right\rangle $ and the $\left\langle 100\right\rangle $
directions to the GPAW results.}
\end{figure}

 We use supervised learning as implemented in PyTorch \citep{PyTorch}
to train the ANNs, where the DFT calculations are employed to obtain
the training data set. Based on the 2D monolayer SnTe, which has the
lattice constant $a_{0}=6.1836$\,\AA and a vacuum layer of $4a_{0}$
along the $z$ direction as shown in Fig. \ref{fig:1_toplogy} (a),
\emph{ab initio} molecular dynamics simulations with GPAW \citep{GPAW}
are performed to simulate a $2\times2$ system (16 atoms) from 500\,K
to 0\,K to generate training samples. Additional configurations with
random displacements of atoms are also used to train the model so
that it can cope with more complex situations. In these calculations,
GPAW uses plane waves with a cutoff energy of 900\,eV, a $2\times2\times1$
Brillouin-zone sampling grid \citep{K-point}, and the Perdew-Burke-Ernzerhof
(PBE) exchange-correlation functional \citep{PBE}. A total of about
8000 configurations are calculated, of which 7900 was used to optimize
the ANN, and 100 was used as a preliminary test of the predictive
ability of the ANNs. For each of the configuration, its energy is
calculated using the energy of the original configuration $E_{O}$,
where none of the atoms is displaced, as the energy reference. Figure
\ref{fig:error} compares the values predicted by the ANN and calculated
by GPAW, which shows a good agreement where the maximum difference
is within 6\textbf{ }meV per atom. We note that the energies of the
chosen configurations used in Fig.\,\ref{fig:error} are all higher
than $E_{O}$. The prediction by the ANNs for configurations with
lower energies agree even better with GPAW as we discuss below (see
Fig.\,\ref{fig:result}).

To further verify the accuracy of the ANN, we also generated special
configurations where all the Sn atoms are synchronously displaced
along a particular direction to a distance $d$ as shown in Fig.\,\ref{fig:result}(a).
We use the trained ANN to predict the energies of these configurations
and generate the PES of SnTe as shown in Fig.\,\ref{fig:result}(b).
It is interesting to see that the ANN has generated a smooth PES with
multiple local energy minimums. To quantitatively check the predictions
of the ANN, we have sampled along the $\left\langle 110\right\rangle $
and the $\left\langle 100\right\rangle $ directions and compared
the energy predicted by the ANN to those calculated with GPAW in Figs.\,\ref{fig:result}(c)
and (d). For the region of interest ($d\leq0.4$\,\AA), the accuracy
provided by the ANN is stunning. It is worth noting that the ANN has
successfully reproduced the double-well potential which is a critical
indication of possible structural phase transitions\citep{ferroelectricity1992}.
This feat is remarkable when we recall that the ANN has a universal
internal structure and is trained with configurations having essentially
random atom displacements. While Figs.\,\ref{fig:result}(c) and
(d) show that the ANN can fit the PES very well along the $\left\langle 100\right\rangle $
and the $\left\langle 110\right\rangle $ directions, one needs to
keep in mind that the power of the ANN lies in their ability to predict
the energy of \emph{any} configuration, and Figs.\,\ref{fig:result}(b-d)
just show some special cases. 

\begin{figure}[h]
\begin{centering}
\includegraphics[width=7cm]{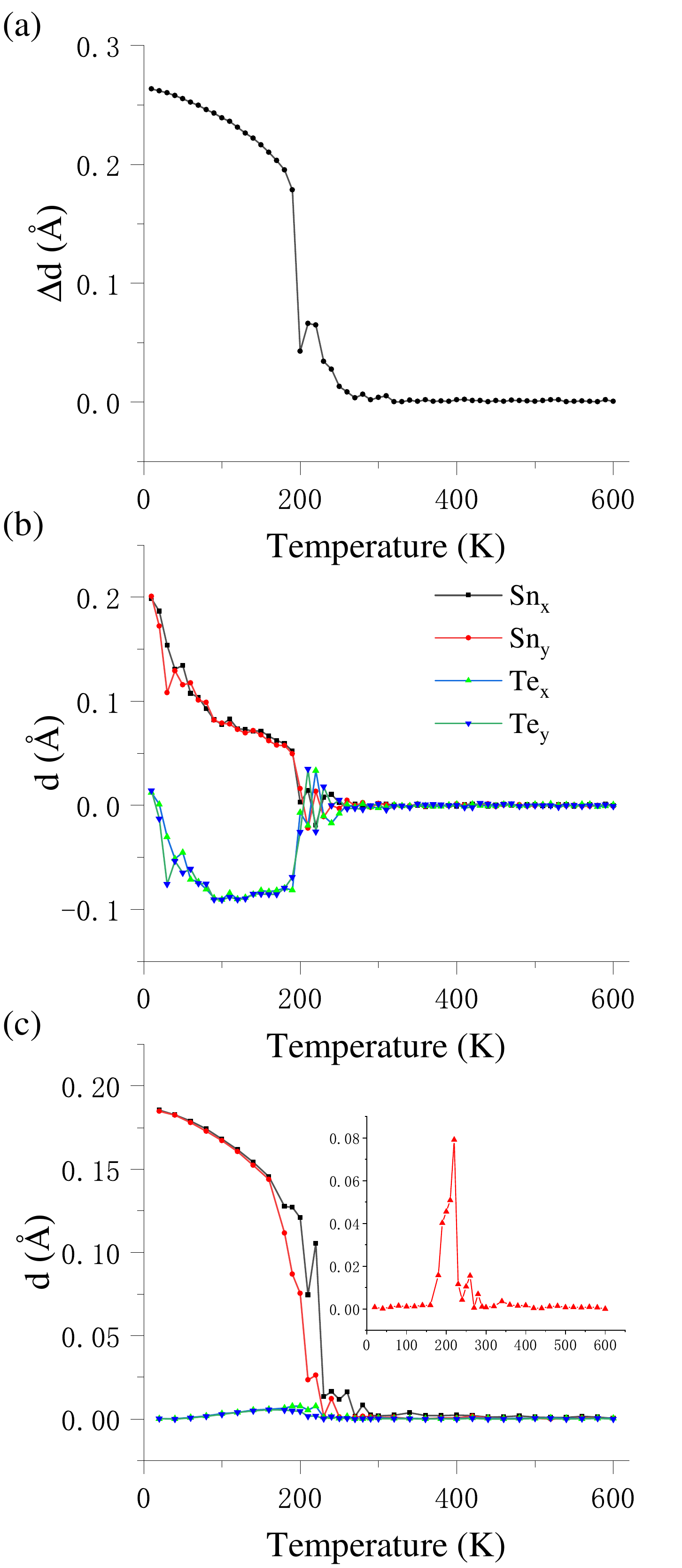}
\par\end{centering}
\caption{\label{fig:Phasetransition} (a) Average relative displacement between
Sn and Te atoms versus temperature (see Eq. (\ref{eq:delta_d}) );
(b) Average positions of Sn and Te atoms versus temperature; (c) Between
160K and 280K, the displacement of Sn atom is no longer exactly along
the $\left\langle 110\right\rangle $direction (see the text). The
inset in (c) shows the displacement difference between the $x$ and
$y$ directions.}
\end{figure}
 Having constructed and trained the ANNs, we now employ them in MC
simulations as energy calculators. The MC simulations use our home-brewed
PyMC$^{2}$, which is a modular Monte-Carlo simulation program specially
designed for crystals \citep{pymc2}. In MC simulations, the ANNs
are used to predicts how much energy change arises when an atom is
displaced. To find the phase transition temperature, we set up a 2D
$12\times12$ supercell (576 atoms), and gradually cool down the system
from 600\,K to 10\,K with a step of 10\,K. At each temperature,
we sweep the systems 80,000 times, in each sweep all atoms are displaced
randomly and such moves are accepted or rejected depending on the
incurred change of energy. Figures\,\ref{fig:Phasetransition}(a)
shows the temperature evolution of the average relative displacement
$\Delta d$, which is determined by the displacements of Sn and Te
within the same unit cell, \emph{i.e.}, 
\begin{equation}
\Delta d=\frac{1}{N}\sum_{i=1}^{N}\left\langle \left|\boldsymbol{d}_{\textrm{Te}}-\boldsymbol{d}_{\textrm{Sn}}\right|\right\rangle ,\label{eq:delta_d}
\end{equation}
where $\left\langle \dots\right\rangle $ indicates the supercell
average, and $N$ is the number of MC sweeps used for the final average.
As we can see from Fig.\,\ref{fig:Phasetransition}(a), a phase transition
occurs at around 260\,K, where the average displacement starts to
increase with the decreasing temperature, gradually reaching a value
around 0.26\,\AA. Figure \ref{fig:Phasetransition}(b) shows that
the averaged displacement of Sn and Te separately, indicating that
the relative displacements are along the $\left\langle 110\right\rangle $
direction. These results are consistent with the energy minimum at
$d=0.26$\,\AA~as shown in Fig. \ref{fig:result}(c). They also
indicate that the monolayer SnTe has a comparable phase transition
temperature as the 1UC SnTe which are previously investigated both
experimentally \citep{SnTe} and theoretically \citep{Qi-JunYe}.

Figures\,\ref{fig:Phasetransition}(a) and (b) also show that the
Sn and Te atoms fluctuate violently around 200\,K, which is likely
due to the competition between the energy minimums (see Fig.\,\ref{fig:result}).
In order to understand the effects of the multiple energy minimums,
we also simulated the system with slightly different setups. For Figs.\,\ref{fig:Phasetransition}(a)
and (b), only the Te atom at the origin is fixed; for Fig. \ref{fig:Phasetransition}(c),
we mimic the epitaxially strained SnTe structure by fixing the Te
atoms at the corner of each unit cell (there are two Te atoms in each
unit cell and one is fixed) so that the crystal lattice structure
is fixed. As shown in Fig. \ref{fig:Phasetransition}(c), there is
a narrow temperature range (between 160 and 280K) where the displacement
of Sn atom is no longer along the direction of $\left\langle 110\right\rangle $,
but more in the direction of $\left\langle 100\right\rangle $. The
snapshots of atom displacements indicate that, in this temperature
range, there are Sn atoms shift along different $\left\langle 110\right\rangle $
directions, resulting in average displacement along the $\left\langle 100\right\rangle $
direction. Therefore, a second ferroelectric phase can be expected
in the phase transition of monolayer SnTe, albeit it may only exist
in a very narrow temperature range.

\begin{figure}[h]
\begin{centering}
\includegraphics[width=8cm]{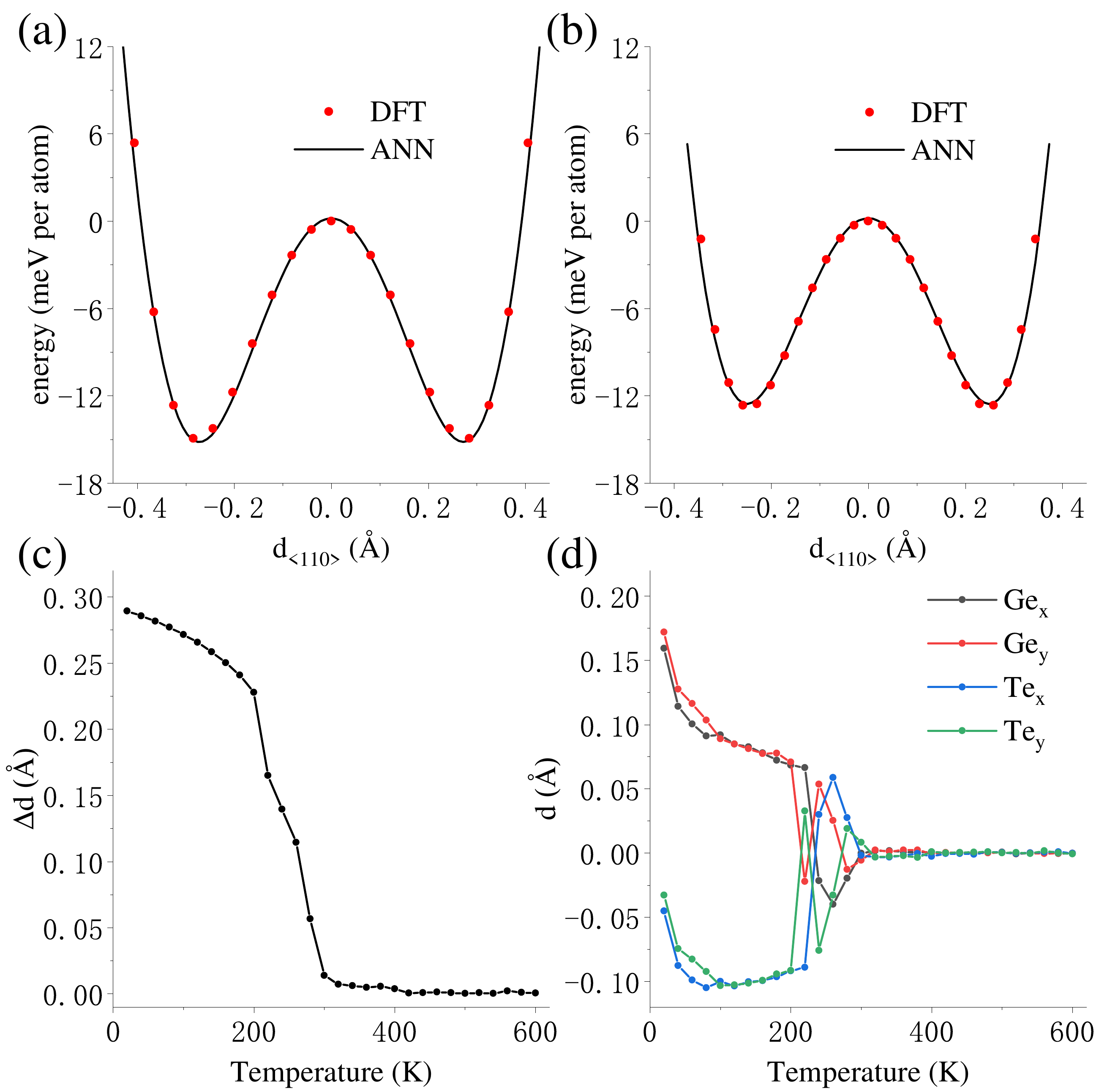}
\par\end{centering}
\caption{\label{fig:GeTe}(a) and (b) shows the ANN's prediction results for
GeTe; (c) Average relative displacement between Ge and Te atoms versus
temperature; (d) Average position of Ge and Te atoms versus temperature.}
\end{figure}
 As mentioned before, the proposed approach can be transferred to
a similar system, such as 2D monolayer GeTe. With the same procedure,
we trained and tested ANNs for GeTe, which show excellent predictive
abilities {[}see Figs. \ref{fig:GeTe}(a) and (b){]}. With the ANNs,
we again conducted MC simulations for monolayer GeTe, and found that
GeTe, similar to SnTe, also have a structural phase transition with
a phase transition temperature around 300\,K as shown in Figs.\,\ref{fig:GeTe}
(c) and (d). The phase transition temperature is higher than that
of SnTe, consistent with the deeper potential well of GeTe.

In summary, we have adopted a ML approach to build and train ANNs
that are employed in MC simulations to investigate the structural
phase transition of ferroelectric monolayer SnTe and GeTe. Unlike
other approaches, no concrete model or formula, which often requires
\emph{a priori} knowledge or a good understanding of the given system,
is necessary to approximate the PES, therefore also removing the difficulty
in determining the coefficients, essentially resulting in a model-less
approach while achieving accurate prediction of their phase transition
temperature. It is demonstrated that the ANNs can work as a special
universal mathematical structure that are capable for various systems
as long as the training data from \emph{ab initio} computation are
available. Such virtues make the ML approach very general and flexible
to investigate structural phase transitions of various systems.
\begin{acknowledgments}
This work is financially supported by the National Natural Science
Foundation of China, Grant Nos. 11974268 and 12111530061. X.C. thanks
the financial support from Academy of Finland Projects 308647. X.C.
and D.W. thank the support form CSC (IT Center for Science, Finland),
project 2001447, for providing computation resources. D.W. also thanks
the support from the Chinese Scholarship Council (201706285020).
\end{acknowledgments}

\end{document}